\documentclass[a4paper]{jpconf}
\usepackage{amssymb,cite,epsfig}
\usepackage{amsthm}
\usepackage{latexsym}
\def\L{{\cal{L}}}

\def\a{\alpha}
\def\b{\beta}
\def\g{\gamma}

\def\d{\delta}

\def\La{\Lambda}
\def\k{\kappa}
\def\m{\mu}
\def\n{\nu}

\def\ep{\epsilon}
\def\th{\theta}

\newcommand{\be}{\begin{equation}}
\newcommand{\ee}{\end{equation}}
\newcommand{\bea}{\begin{eqnarray}}
\newcommand{\eea}{\end{eqnarray}}
\newcommand{\beqar}{\begin{eqnarray*}}
\newcommand{\eeqar}{\end{eqnarray*}}

\begin{document}

\title{\textbf{Properties of codimension-2 braneworlds \\ in six-dimensional Lovelock theory}}

\author{\textbf{Christos Charmousis}$^{1,2}$}
\address{$^1$Laboratoire de Physique Th\'eorique,\\ Universit\'e de Paris-Sud,\\
B\^at. 210, 91405 Orsay CEDEX, France\\
$^2$Laboratoire de Mathematiques et Physique Th\'eorique,\\ Universit\'e Francois Rabelais,\\
Parc de Grandmount, 37200 Tours, France}
\ead{\textbf{christos.charmousis@th.u-psud.fr}}

\author{\textbf{Antonios Papazoglou}}
\address{Institute of Cosmology and Gravitation,\\ University of
Portsmouth, \\Portsmouth PO1 3FX, UK }
\ead{\textbf{antonios.papazoglou@port.ac.uk}}

\begin{abstract}
We consider maximally symmetric 3-branes
embedded in a  six-dimensional bulk spacetime with Lovelock
 dynamics. We study the properties of the solutions with
respect to their induced curvature, their vacuum energy and their
effective compactness in the extra dimensions. Some simple
solutions are shown to give rise to self-accelerating braneworlds,
whereas several others solutions have self-tuning properties. For the case of
geometric self-acceleration we argue that the cross-over scale in between four-dimensional and
higher-dimensional gravity and the scale of late-time geometric acceleration,
fixed by the present horizon size, are  related via the conical deficit angle
of the six-dimensional bulk solution, which is a free parameter.
 \end{abstract}

\section{Introduction}

Cosmological  and astrophysical data indicate that more than two
thirds of the content of the Universe is of the form of dark
energy. The  best fit to the data indicates that this component
has the form of a small cosmological constant.  This results to a
 significant theoretical problem since its natural value, from the point
 of view of Quantum Field
Theory, is of the order of the ultraviolet cutoff we would impose
for our quantum field theory (anything from SUSY breaking scale to
the Planck scale). In addition, one has to understand why this
cosmological constant is of the order of the dark matter energy
density now, taking into account the completely different
cosmological evolution of these components of the Universe.

The above, of course, are correct statements given that we have a
homogeneous Universe described by the Einstein's field equations.
 Taking into account the above difficulties, modifying gravity in the infrared
is a legitimate theoretical hypothesis that should be taken
seriously. In the context of braneworld models, there have been
some interesting ideas. Firstly, it was shown that five-dimensional models with an
induced gravity term on the brane (DGP models) can have a {\it
self-accelerating} phase \cite{deffayet}, where the current acceleration of the
Universe is not due to a compontent of the Universe energy
density, but rather a geometrical effect. An even more ambitious
proposal, that of the {\it self-tuning} \cite{5dself,6dself}, was to find models where
the vacuum energy of the brane can be large without affecting the
curvature of the brane and without fine-tuning of it with other
brane or bulk parameters. Both proposals change the way that we
see the interplay between vacuum energy and curvature in gravity.
There were not, however, without problems. Self-accelerating
braneworlds were typically infested by ghosts \cite{kal} and self-tuning
ones had hidden fine-tunings or curvature singularities \cite{sing}.

In our recent work \cite{charpap}, we examined a  completely novel
possibility of obtaining acceleration due to geometry as well as
certain self-tuning properties. The modified gravity theory that
we will study is Lovelock theory \cite{Lovelockth} in six dimensions, which is the
natural extension of General Relativity in higher dimensions. The
Lovelock theory in six dimensions has, in addition to the
Einstein-Hilbert term, the Gauss-Bonnet combination (for a recent review on Lovelock theory see \cite{love}).
Although the
latter is a topological invariant in four dimensions, it becomes
dynamical for higher dimensions and modifies the gravitational
theory. This theory is special because, on the one hand
 it provides geometric novel solutions which are absent in Einstein theory,
and on the other hand, it gives  an induced gravity term on the
brane.  In the context of this theory, we found examples of both
self-accelerating and self-tuning cases. These examples open new
possibilities for consistent self-acceleration and effective
self-tuning which need to be considered in more detail in the
future.

\section{Lovelock braneworlds}

Let us consider the six-dimensional dynamics of Lovelock gravity
 with a bare cosmological constant $\La$ and a Gauss-Bonnet
term. The action of the system reads \be \label{chaaction} S=\int
d^6 x\, \sqrt{-g}\left[\frac{1}{16 \pi G_6}(R +\hat{\alpha}
\L_{GB})-2\La\right]  ~, \ee where \be\L_{GB}=R_{MNK \La}R^{MNK
\La} -4R_{MN}R^{MN}+R^2 ~, \label{GBaction} \ee is the
Gauss-Bonnet Langrangian density, $G_6$ the six-dimensional
Newton's constant and $\hat{\a}$ the Gauss-Bonnet coupling.

The procedure we use to generate brane world solutions, is to
doubly Wick rotate black hole solutions (first studied by Boulware and Deser \cite{boulware}, see also \cite{cai}). In this procedure, the
positions of the horizons $r_h$ will be the endpoints of the
internal space where there codimension-2 branes are in principle
located. The brane tension is in fact related to the temperature of the black hole horizon, and a warped two brane setup will correspond to black hole solutions with double horizons. In fact, via a generalised version of Birkhoff's theorem \cite{zegers}, it can be
shown that the most general axisymmetric solutions of
(\ref{chaaction}) with maximally symmetric four-dimensional
subspaces are \be \label{chasol} ds^2=V(r)
d\theta^2+\frac{dr^2}{V(r)}+r^2 h^{(\k)}_{\m\n}dx^\m dx^\n~, \ee
where the potential is given by \be \label{chapot} V(r)=\kappa +
\frac{r^2}{2\alpha}\left[1+\epsilon \sqrt{1+4\alpha
\left(a^2-\frac{\ep \mu}{r^{5}}\right)}\right]~, \ee with
$\alpha=6\hat{\alpha}$, $16 \pi G_6 \Lambda=20a^2$ the positive cosmological constant{\footnote{Here we have omitted the bulk charge parameter for simplicity.}}. The
four-dimensional metric brane metric $h^{(\k)}_{\m\n}$ is  parametrised by
$\kappa=0,-1,1$, for four-dimensional Minkowski, $AdS_4$ and
$dS_4$ respectively with curvature $R[h]=12\k$. Finally,
$\epsilon=\pm 1$, giving rise to two distinct branches of
solutions. The branch $\ep=+1$ ({\it Gauss-Bonnet branch}) does
not have an Einstein theory limit as $\a \rightarrow 0$ (more recently the vacuum in this branch was shown to be unstable \cite{tony}). This
limit is regular for the other branch with $\ep=-1$ ({\it Einstein
branch}). The case where $1+4\a a^2=0$ is special, because the
theory can be written in a Born-Infeld (BI) form \cite{zanelli}. It does not have
an Einstein theory limit, however, this is the only case that we
have a unique vacuum.

Defining the Gaussian Normal radial coordinate
$\rho=\sqrt{4(r-r_h)/V'_{r_h}}$ and expanding around e.g. one root
$r_h$ of $V$, we get that the internal space is locally conical
\be ds_2^2\approx \left({1 \over 4} V_{r_h}'^2\right) \rho^2
d\theta^2+ d\rho^2~. \ee If the angular coordinate has periodicity
$\th \in [0,2\pi c)$, then the deficit angle which is induced at
the brane position is $\d = 2 \pi (1 - \b)$ with $ \b = {1 \over
2}|V'_h|c$. Note that the conical deficit is related as usual
to the temperature of the Wick rotated black hole horizon.
From the Lovelock equations supplemented by a brane
tension term, one can separate the distributional Dirac parts and
write down induced Einstein equations for the brane. These brane
junction conditions are \cite{z}
\be 2\pi (1-\beta)\left(-\g_{\m \n}
+4\hat{\alpha} G_{\m \n} ^{ind} \right)=8 \pi G_6 T^{brane}_{\m
\n}~, \label{antInduced}\ee where $\g_{\m\n}= r_h^2
h^{(\k)}_{\m\n}$ is the induced metric on the brane with curvature
$R[\g]=12\k/r^2_h \equiv 12\k H^2$, and $G_{\m \n} ^{ind}=-3\k H^2
\gamma_{\mu\nu}$ is the induced Einstein tensor. The brane position $V(r_h)=0$ depends
on the bulk parameters via (\ref{chapot}). The induced
Newton's constant on the brane can be determined from
(\ref{antInduced}) to be \be G_4={3G_6 \over 4\pi \a(1-\b)}~.
\label{antNewt}\ee Note that in order to have positive induced
Newton's constant, we should have have angle deficit ($\b<1$) for
$\a>0$ and angle excess ($\b>1$) for $\a<0$.

Substituting the $G_{\m \n} ^{ind}$ back in (\ref{antInduced}), we
find a relation between the  Hubble parameter $H$ on the brane and
the action parameters ($T^{brane~\n}_\m=-T \d_\m^\n$) \be \k H^2 =
-{1 \over 2 \a} + {8\pi G_4 \over 3} T~. \label{antcontrib}\ee The
above equations (\ref{antNewt}), (\ref{antcontrib}) are very important since they firstly relate the hierarchy in between the scales of the theory and secondly the curvature on the
brane $H^2$ to its sources, namely the brane tension and the
Gauss-Bonnet coupling. We see in particular from  (\ref{antcontrib}) that the junction conditions tell us
that the effective expansion $H$ is in one part due to the
Gauss-Bonnet induced cosmological term and in another  part due to
the vacuum energy of the brane. On the other hand we expect (\ref{antNewt}) to be indicating a cross-over scale in between a four-dimensional gravity phase and a six-dimensional one, $r_c^2 =G_6/G_4$. Unlike the codimension-1 DGP model note the appearance of two scales $\alpha$ and $\beta$ dictating the size of $r_c$.

\section{Self-properties of the solutions}

Let us now discuss the physical consequences of the above
solutions. In particular, we wish to see whether we can obtain
codimension-2 braneworlds  exhibiting  {\it self-accelerating} or
{\it self-tuning} behaviour. The key relation for picking these
solutions is (\ref{antcontrib}).

\subsection{Self-acceleration}

For self-acceleration we need $dS_4$ vacua ($\k=1$), where one has
a positive geometrical contribution to the curvature, i.e. $\a<0$
coming from the Gauss-Bonnet term in the action (\ref{chaaction}).
In addition, this contribution should be dominant in comparison to
the brane tension contribution (in other words the $T$-term in
(\ref{antcontrib}) should be negligible). There are two cases
where this can happen.

First, if we have no bulk cosmological constant $a^2=0$ and we are
in the Einstein branch $\ep=-1$. Then the internal space is
non-compact and the brane position is bounded from below as
$r_h(\m)>\sqrt{2|\a|}$. The limit $r_h(\m_s)=\sqrt{2|\a|}$, with
$\m_s \equiv \sqrt{2} |\a|^{3/2}$, is singular since it
corresponds to the branch cut singularity of the square root of
(\ref{chapot}). The solution is self-accelerating when $\m$ is in
the neighborhood of $\m_s$. The bulk  solution is simply a Wick
rotation of the six-dimensional Schwarzschild black hole.

Second, if the mass parameter vanishes $\m=0$ and we are in the
neighborhood of the BI point $a_{BI}^2=1/(4|\a|)$ (in both
branches $\ep=\pm 1$). Then, the brane position is a function of
the bulk cosmological constant $r_h=r_h(a^2)$ and at the BI limit
$r_h(a^2_{BI})=\sqrt{2|\a|}$. In these cases the internal spaces
are compact and since the space has no singularity at $r=0$, we
can extend the radial coordinate to $r<0$ and consider the region
of $-r_h \leq r \leq r_h$. The internal space is symmetric around
$r=0$, thus we have $Z_2$ symmetry around the equator of the
internal space.

In both the above cases in order for the geometrical
acceleration to account for the current acceleration of the
Universe, the Gauss-Bonnet coupling appearing in the 6 dimensional action should be enormous, roughly of
the order $\a \sim 10^{120} M_{Pl}^{-2}$. This hierarchy means that the bulk gravity is essentially dictated by the higher
order Lovelock term that in turn gives ordinary four-dimensional gravity on the brane according to (\ref{antInduced}). The hierarchy in between $G_4$ and $G_6$ maybe reduced by sufficiently fine-tuning $\beta$ to be close to $1^-$.  In fact combining  (\ref{antNewt}) and (\ref{antcontrib}), for $T \approx 0$, we have that,
\be
 \delta \sim \frac{3 r_c^2}{2 r_0^2}
\ee
where $r_0=H_0^{-1}$ is the horizon size and $\delta$ the angular excess, $\delta<0$.
Hence the crossover scale, $r_c$, is now a combination of a purely topological number $\delta$ and the horizon size $r_0$, unlike the situation in codimension-1 DGP where the two scales are the same. We should emphasize however, that strictly speaking the cross-over scale should be obtained from the brane propagator of the above bulk solution (\ref{chasol}) and is not necessarilly the scale appearing in the junction condition. Clearly this requires further investigation.

\subsection{Self-tuning}

For self-tuning to operate, one should be able to absorb
variations of the brane tension in integration constants like $c$
and $\m$, with the crucial demand of keeping the curvature of that
brane as well as other bulk or brane parameters constant. We will
be obviously interested in  $dS_4$ ($\k=1$) or flat ($\k=0$)
vacua. It turns out that if there is more than one brane present
in the compactification, there are unavoidable fine-tunings in the
model. Thus, the only possible self-tuning vacua can be the ones
with only one brane present, or when an extra mirror brane is
present, as in the $Z_2$-symmetric model that we mentioned in the
previous subsection. A supplementary requirement for the
self-tuning to be satisfying, is that the scales of $H^2$ and $T$
should be dissociated. For the latter to happen, one should have
$|\a|H^2 \ll 1$ (in other words the $H$-term in (\ref{antcontrib})
should be negligible). In this case we can see from
(\ref{antcontrib}) and (\ref{antNewt}) that changes to $T$ can be
absorbed in $c$. Moreover, $\a$ can have much more natural values, i.e. $|\a| \sim M_{Pl}^{-2}$,  than
the ones for the previous self-accelerating vacua. In all these cases,
since $c$ should follow the variation of $T$, we will obtain a
vacuum-energy-dependent Newton's constant $G_4$ as seen from
(\ref{antNewt}). There are several examples where this can happen.

First, all the non-compact flat vacua, for $\a>0$ or $\a<0$, and
with or without bulk cosmological constant,  satisfy the above
requirements. In this cases, we have exact self-tuning solutions.

For the $dS_4$ vacua, there are many possibilities with
non-compact or compact internal space. A non-compact example  is
when we have no bulk cosmological constant $a^2=0$, we are in the
Einstein branch $\ep=-1$, for both $\a>0$ and $\a<0$, and for
large enough mass $\m |\a|^{-3/2} \gg 1$. On the other hand, the
only  compact ($Z_2$-symmetric) case is when the mass parameter
vanishes $\m=0$, we are in the Einstein branch $\ep=-1$, for
$\a<0$ and for small enough bulk cosmological constant $|\a| a^2
\ll 1$.

\section{Discussion and Conclusions}

We have obtained several self-accelerating and self-tuning
solutions in a Lovelock six-dimensional theory with codimension-2
branes. The new setting of these models may help overcome the
obstacles of the previous self-accelerating and self-tuning
 models in the literature.

An important property of the above models is that the
codimension-2 junction conditions have an induced  Einstein tensor
term. Therefore, one should expect that the theory (even for
non-compact or very large volume models) behaves in a
four-dimensional way upto some cross-over scale which we have argued to be
$r_c^2=G_6/G_4$.
We should emphasise that a characteristic element of the present
setting is that this cross-over scale is in principle different
from the scale of self-acceleration $r_0$ because the deficit angle
$\delta$ which comes also into play. In principle the scale $r_c$
could be much lower than the horizon size $r_0$ thus reducing the hierarchy between $G_6$ and
$G_4$. How low we can set the crossover scale $r_c$, compared to $r_0$
depends on the modified higher dimensional gravity
 and how much it will modify large scale cosmological observables.

What is, therefore, the next step in the analysis of the codimension-2 brane models presented here, is the linear perturbation of
these solutions as well as the cosmology analysis. The former, will tell us of the stability and the
precise gravitational spectrum. For the latter, although one
expects that the introduction of matter on the brane introduces
singularities which need to be regularised, the fact that there
exists an induced Einstein equation on the brane, may allow for
the cosmology to be studied without the need of an explicit
regularisation. Furthermore, knowledge of the modified Friedmann equations on the brane
will tell us a lot more on the cosmological scales that we have hinted upon here and on the validity of
codimension-2 braneworlds.

\ack A.P. is supported by the Marie Curie
Intra-European Fellowship EIF-039189.

\end{document}